\documentclass[conference,letter]{IEEEtran}

\usepackage{authblk}
\usepackage{tabularx}
\usepackage{booktabs}
\usepackage{mathtools}
\usepackage{tikz}
\usepackage{algorithm}
\usepackage{multirow}
\usepackage{algpseudocode}  
\usepackage[T1]{fontenc}
\usepackage{array}
\usepackage{booktabs}
\usepackage{latexsym}
\usepackage{caption}
\usepackage{subcaption}

\usepackage{enumitem}
\usepackage{url}
\graphicspath{{./figures/}}
\usepackage{times}

\PassOptionsToPackage{hyphens}{url}\usepackage{hyperref}

\begin{document}

% Example definitions.
% --------------------
\def\x{{\mathbf x}}
\def\L{{\cal L}}

% Title.
% ------

\title{Understanding the Smartrouter-based Peer CDN for Video Streaming}

%
% Single address.
% ---------------

%\address{}

\author[1]{Ming Ma}
\author[2]{Zhi Wang}
\author[1]{Ke Su}
\author[1]{Lifeng Sun}
\affil[1]{Tsinghua National Laboratory for Information Science and Technology \authorcr Department of Computer Science and Technology, Tsinghua University }
\affil[2]{Graduate School at Shenzhen, Tsinghua University}
\affil[ ]{\{mm13@mails., wangzhi@sz., suk14@mails., sunlf@\}tsinghua.edu.cn}

\maketitle

%On crowdsourced interactive live streaming: a Twitch. tv-based measurement study
\begin{abstract}

	Recent years have witnessed a new video delivery paradigm: smartrouter-based video delivery network, which is enabled by smartrouters deployed at users' homes, together with the conventional video servers deployed in the datacenters. Recently, ChinaCache, a large content delivery network (CDN) provider, and Youku, a video service provider using smartrouters to assist video delivery, announced their cooperation to create a new paradigm of content delivery based on householders' network resources \cite{youkuchinacache}. This new paradigm is different from the conventional peer-to-peer (P2P) approach, because such dedicated smartrouters are inherently operated by the centralized video service providers in a coordinative manner. It is intriguing to study the strategies, performance and potential impact on the content delivery ecosystem of such peer CDN systems. In this paper, we study the Youku peer CDN, which has deployed over $300$K smartrouter devices for its video streaming. In our measurement, $78$K videos were investigated and $3$TB traffic has been analyzed, over controlled routers and players. Our contributions are the following measurement insights. First, a global replication and caching strategy is essential for the peer CDN systems, and proactively scheduling replication and caching on a daily basis can guarantee their performance. Second, such peer CDN deployment can itself form an effective Quality of Service (QoS) monitoring sub-system, which can be used for fine-grained user request redirection. We also provide our analysis on the performance issues and potential improvements to the peer CDN systems.
	
\end{abstract}

\section{Introduction} \label{section:introduction}

To meet the skyrocketing growth of bandwidth requirement for data-intensive video streaming, and reduce the monetary cost for renting expansive resources in conventional content delivery networks (CDNs), video service providers today are deploying their \emph{peer CDNs} to make use of network and storage resources at individuals' homes for content delivery. Youku, a largest online video provider, has deployed over $300$K smartrouters at their users' homes in less than one year \cite{youku1million}, expecting to turn a large fraction of its users ($250$M) of Video-on-Demand (VoD) to such content delivery peer nodes. 

\begin{figure}[!t]
	\centering
		\includegraphics[width=\linewidth]{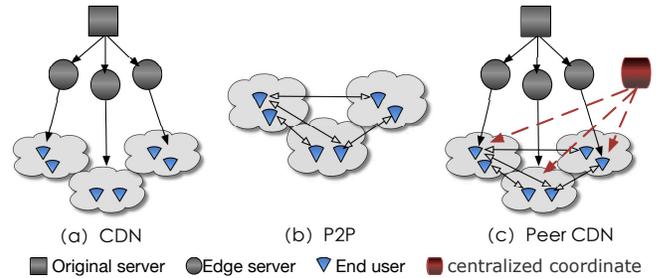}
		\caption{The system architecture for CDN, P2P, and peer CDN.}
		%\vspace{-0.3cm}
	\label{fig:CDN_P2P_PCDN}
	\vspace{-0.1cm}
\end{figure}

In Fig.~\ref{fig:CDN_P2P_PCDN}, we plot the content delivery paradigms including conventional CDN, peer-to-peer (P2P) and peer CDN. Compared to the conventional CDN approach, peer CDN employs network resources that are much closer to users, e.g., it can serve as much as $80\%$ of the content requests by peer nodes in hundreds of meters; While compared to the conventional P2P paradigm where users individually cache and serve each other, the nodes in a smartrouter-based peer CDN are closely coordinated by the centralized knowledge. For example, the content provider schedules nodes in a peer CDN to proactively cache contents, and redirects users to download content from particular smartrouters. Today, content providers and such peer CDN providers even start collaborating to change the traditional content delivery paradigm, to satisfy the ever increasing generated content and edge network requests \cite{youkuchinacache}. 

As such content delivery paradigm can fundamentally change the deployment of content services and applications, and even the roles of individuals play in the Internet, e.g., any individual can become not only a content publisher, but also a corresponding content source hosting the service for the contents generated by herself, it is thus intriguing to investigate the details of the peer CDN paradigm, including its strategies, performance, and limitations.

In this paper, we conduct extensive measurement to study video streaming by its peer CDN in a largest video service provider, Youku, which has attracted over $300$K users to deploy the smartrouters (namely Youku Router) with $8$GB storage at their homes/offices to serve Youku users. The challenges to study the system performance and key strategies are as follows: First, the system is distributed in nature, it is difficult to have a global knowledge using only limited scale of active experiments; Second, the system is a combination of both conventional CDN servers and peering nodes, it is challenging to identify the target peer nodes we are interested in; Third, incentive mechanisms are deployed in the system, which leads to even more noise in our traffic analysis.

To address these challenges, we design \emph{active} and \emph{passive} measurement experiments to study not only its architecture, but also key strategies. On one hand, we deploy controlled clients in different networks (e.g., CERNET, Unicom, etc.) to interact with the peer CDN nodes, and study their behaviors; On the other hand, we deploy controlled smartrouters in different cities and ISPs to passively observe how these nodes are controlled by the whole system, and how they serve others. Our contributions are summarized as follows.

\begin{itemize}
\item  Based on our measurement studies covering $78$K videos, $126$ conventional centralized CDN servers and $3$M users, we present not only the architecture and protocols, but also the key strategies that affect the performance of a joint CDN and peer CDN system.
			 
\item  We reveal that the global content replication, replacement and user redirection strategies are essential for the peer CDN paradigm, and proactively scheduling replication and caching on an hourly basis can guarantee the performance of the system. To name a few results: (1) Smartrouters use a frequently content update strategy to meet the user demand, e.g., the median lifespan of chunks cached in the smartrouters is $24.2$ hours; (2) Using global knowledge, smartrouters are scheduled by a centralized peer selection mechanism, e.g., based on ISP or location.

\item We examine the system performance and observe that: (1) The peer CDN can guarantee the system Quality of Service (QoS), both for the start-up delay and the download speed; (2) Youku peer CDN fully utilizes the smartrouter resources of edge network, e.g., $80\%$ of the content requests can be served with at least $70$\% of the data which come from smartrouters. We also propose some possible strategies to improve the system performance for the future development.
\end{itemize}

The rest of the paper is organized as follows. In Sec.~\ref{sec:measure}, we introduce our measurement scheme. The video peer CDN architecture and system workflows are presented in Sec.~\ref{section:architecture}. The strategies of video placement and peer selection are analyzed in Sec.~\ref{section:strategy}. We evaluate the system performance in Sec.~\ref{section:qos}. We also present our discussion in Sec.~\ref{sec:discussion}. Finally we analyze related works in Sec.~\ref{section:relatedwork} and conclude this paper in Sec.~\ref{section:conclusion}.

\section{Measurement Methodology}  \label{sec:measure}

We conduct both passive and active measurement on the peer CDN nodes (i.e., Youku smartrouters) and clients in our study. 
\subsection{Measurement by Controlled Peer CDN Nodes}
\subsubsection{Testing Node Deployment}

\begin{figure}[!t]		
	\centering
	\includegraphics[width=0.8\linewidth]{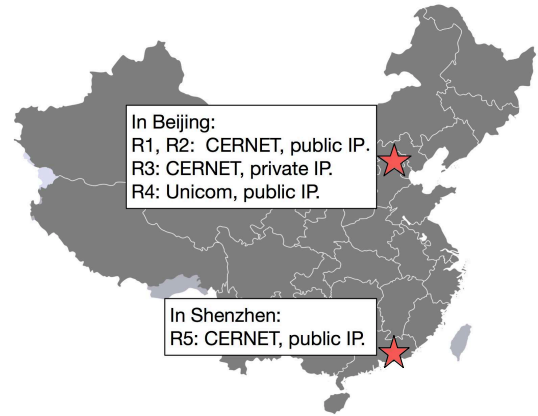}
	\caption{Deployment of testing peer CDN nodes.}
	\label{fig:nodelocation}
		 \vspace{-0.1cm}
\end{figure}

 %$\rhd$ Measurement by Controlled Peer CDN Nodes. 
  In order to ``sensing'' the strategies used in Youku peer CDN, we deploy $5$ Youku smartrouters in different locations with different ISPs in our measurement study, and observe the interactions between these smartrouters and other CDN nodes. As illustrated in Fig.~\ref{fig:nodelocation}, the $5$ routers are denoted as follows: 
 \begin{itemize}
 \item \textsf{R1},\textsf{R2}, which are deployed in Beijing (ISP: CERNET), with public IP address and $80$Mbps downlink/uplink bandwidth; 
 \item \textsf{R3}, which is deployed in Beijing (ISP: CERNET), with private IP, based on Network Address Translator (NAT), and $80$Mbps downlink/uplink bandwidth; 
 \item \textsf{R4}, which is deployed in Beijing (ISP: China Unicom), with public IP and $4$Mbps downlink bandwidth and $512$Kbps uplink bandwidth;  
 \item \textsf{R5}, which is deployed in Shenzhen (ISP: CERNET), with public IP address and $80$Mbps downlink/uplink bandwidth.

 \end{itemize}
 
\subsubsection{Device Information}
We use the ordinary version of Youku smartrouters as our testing nodes, and the hardware information is illustrated in Table \ref{tab:smartroutercost}. In particular,  each device provides a $580$MHz CPU, and the capacity of the internal storage is one $8$GB TF cards.

\begin{table}[!t]
  \caption{Hardware information of controlled testing nodes. }
  \label{tab:smartroutercost}
  \centering  
   %\begin{threeparttable}
  \begin{tabular}{ll}%{ *3{>{\centering\arraybackslash}m{1.5 in}} @{}m{0pt}@{}}
    \toprule
        Parameter	&	Information \\
    \midrule
        CPU				&	$580$MHz	\\
        RAM				&	$128$MB (DDR2)	\\
	    Operating System	&	OpenWrt $2.6.36$	\\
	    WiFi Protocol/Channel   & IEEE $802.11$ b/g/n@$2.4$ GHz	\\
	     Storage &  $8$GB TF cards	\\
         Price		&	$26.3$USD (December 2015)	\\    
     \bottomrule
  \end{tabular}
   \vspace{-0.1cm}
\end{table}

\begin{table}[!t] %\footnotesize
  \caption{Statistics in our measurement studies.}
  \label{tab:datasets}
  \centering  
    \begin{tabular}{ll}%{p{.65\linewidth}p{.35\linewidth}}
    \toprule
     Time period		&	9/15  -- 10/30, 2015	\\
    \midrule 
     Video number	    & $78,285$ \\
     Total traffic 	& $3.1$TB \\
     TCP/UDP Packets				&	$4,438,510,988$	\\
	 Contacted servers	    &	$126$		\\
	 Contacted IPs				&  	$3,155,877$\\
	 Distinct ASes			&  $4,015$\\
    \bottomrule
  \end{tabular}
  \vspace{-0.2cm}
\end{table}

\subsubsection{Measurement on the Testing Routers}

There are two types of measurements on the devices:
\begin{itemize}
\item File System Monitoring:
	By performing root injection\footnote{http://openwrt.io/docs/youku/}, we are able to login these devices via \textsf{SSH}, and monitor the folders on the devices where \emph{video files} (chunks and video files are used exchangeably in this paper) are stored to serve users. In Youku peer CDN, all contents are cached as video chunks with an unique content ID, and our experiments cover $78$K different videos during the monitoring period (from September 15th to October 30th, 2015). In Sec.~\ref{section:strategy}, we will present the details of the file monitoring results. 

\item Traffic Monitoring:
 We monitor the traffic patterns on these devices, using the conventional network utility tools including \textsf{tcpdump}, \textsf{netstat}, etc. Through this monitor, we explore the interaction protocols between the smartrouters and the peer CDN servers (and common users). Table \ref{tab:datasets} illustrates the statistics in our measurement experiments. For instance, the dataset contains $126$ unique Youku peer CDN servers and $3$M unique IPs.

\end{itemize}

 \subsection{Measurement by Controlled testing users}

 In our study, we actively run controlled testing users to join the system, and measure their interactions with the peer CDN nodes (including both of our testing routers and others). To be specific, we act as ordinary users who request a number of videos, and monitor:
 \begin{itemize}
 \item How these requests are served by the system, which can reveal the service workflows.
 \item How smartrouters are allocated to cache the videos by analyzing the peer lists returned to the users, which can reveal the content deployment strategies used by Youku.
\end{itemize}  

In the next section, we exhibit the system architecture of the Youku peer CDN inferred from our measurement.

\section{System Architecture} \label{section:architecture}

\begin{table*}[!t]\footnotesize
  \caption{Key HTTP requests captured from our testing routers and users.}
  \label{tab:HTTP}
  \centering  
    \begin{tabular}{p{.08\linewidth}p{.3\linewidth} p{.56\linewidth}}
    \toprule
  Step	&   HTTP message		&				Description	\\
    \midrule  								
  Step \textsf{A} & GET pcdnapi.youku.com/pcdn/sysconf/acc			&Routers request a key for config file downloading.\\
  Step \textsf{A} & GET pcdnapi.youku.com/update/config 			&Routers request the config file to set the operating parameter of the peer router.\\
  Step \textsf{B} & POST pcdnstat.youku.com/iku/log/acc	&Routers report its status to control plane.\\
   Step \textsf{C} & GET /acc/hotspot/cdnurl?rid=$C_{peer}$		 & Routers request to prefetch the chunk $C_{peer}$. \\ 
  Step \textsf{D} \textsf{1}	 & GET /player/getFlvPath/sid/$C_{CDN}$.		&	Routers/Users request Load balance server to obtain the candidate edge CDN server list. \\
  Step \textsf{E}, \textsf{2} & GET /youku/sid/.../$C_{CDN}$.		 & Routers/Users download the chunk from edge servers.  \\
  Step \textsf{3}	& GET /getTaddr?f=$C_{peer}$.	 & Agents get the candidate peer list.  \\
     \bottomrule
  \end{tabular}
  \vspace{-0.2cm}
\end{table*}

	Through our measurement study and protocol analysis, we analyze the workflows of the system and verify the functional roles of each CDN nodes connected by our testing routers and testing users. In this section, we infer the architecture used by the Youku peer CDN system for VoD service.
	
\begin{figure}[!t]
 		\centering
 		\includegraphics[width=0.8\linewidth]{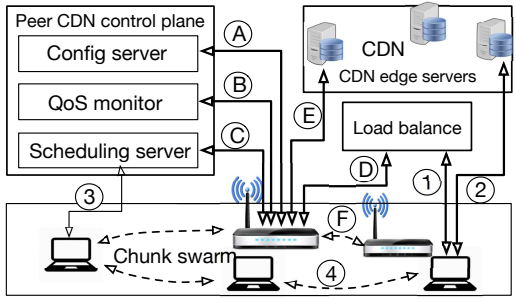}
		\caption{Architecture of the peer CDN system.}
   		\label{fig:architecture}
   		%\vspace{-0.3cm}
\end{figure} 
\begin{figure}[!t]
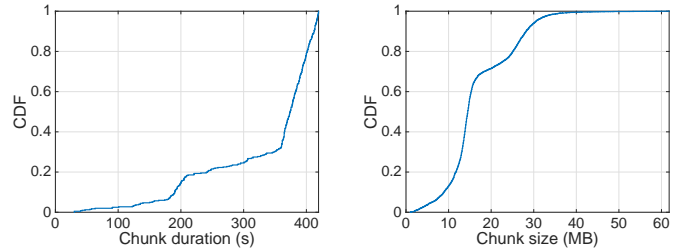

\centering
     \begin{minipage}[t]{\linewidth}
		  \begin{minipage}[t]{.46\linewidth}
			  \centering
				   \includegraphics[width=\linewidth]{Chunk_duration_CDF.eps}
			  \centerline{\scriptsize (a) The CDF of chunk duration.}
		 \end{minipage}
		 \hfill
		 \centering
			\begin{minipage}[t]{.46\linewidth}
			  \centering
				   \includegraphics[width=\linewidth]{Chunk_size_CDF.eps}
			  \centerline{\scriptsize (b) The CDF of chunk size.}
		 \end{minipage}
		 \hfill
     \caption{Statistics of contents replicated by our testing routers.}
     \label{fig:chunkinfo}
   	\end{minipage} 
   	 \vspace{-0.2cm}
\end{figure}

\subsection{Architecture and Workflows}
 Fig.~\ref{fig:architecture} illustrates the general architecture and workflows of the Youku peer CDN system. There are three basic components in Youku peer CDN:
 \begin{itemize}

\item \textbf{Agents}: There are three types of agents in the system, including (1) \emph{Peer routers}: Dedicated smartrouters which can proactively cache and distribute contents; (2) \emph{Peer clients}: Clients who install the Youku Accelerator and cache watched videos to serve others; (3) \emph{Ordinary users}: Users who watch videos on Youku. 

\item \textbf{Peer CDN Controller}: This component decides the primary scheduling strategy, i.e., which videos should be replicated by which peer routers. There are mainly $3$ types of servers as follows:
\begin{itemize}
	\item Config servers: Peer routers download configuration parameters from Config servers (step \textsf{A}). In our datasets, we trace the HTTP response in XML file, e.g., ``<param name=`report-partners-timeval' value=`1800'/>'' indicates that the peer router will report its partners' information every $1800$ seconds.

	\item QoS monitor: Peer routers report statistics (step \textsf{B}) to the QoS monitor, including the information of their partner states and their operation problems, i.e., network congestion and software crash. Thus peer CDN can have a global views of end-to-end QoS using the monitoring mechanism.
		
	\item Scheduling servers: Scheduling servers schedule the content replication (step \textsf{C}) and user redirection (step \textsf{3}) according to the information monitored. Firstly, smartrouters periodically receive replication tasks from the scheduling servers for downloading new contents to their local storage (step \textsf{C}); Secondly, ordinary users discover the candidate peer lists from the scheduling servers (step \textsf{3}).

\end{itemize}

 \item \textbf{CDN Infrastructure}: The CDN infrastructure takes two responsibilities. The first is as a ``back-up'' for the peer routers and users. If peers become unstable, users could download contents from edge servers (step \textsf{1},\textsf{2}); the second is to publish video contents, by pushing the latest content to the peer routers (step \textsf{D},\textsf{E}).
\end{itemize}
	
	Based on the system components illustrated above, we can understanding the cooperation between the Youku CDN infrastructure and the peer CDN control plane. Table \ref{tab:HTTP} illustrates the key HTTP requests issued by our testing routers or testing users. In the Youku peer CDN, the video ID in the CDN server, denoted as $C_{CDN}$ (e.g., ``030008080556377E2A51F503BAF2B1CBC532CF-BA95-0858-9BC4-31009B5D3563''), is different from the video ID of the same chunk in the peer routers, denoted as $C_{peer}$, which is a $40$-byte hex hash value, e.g., ``200000004F6078A3F2D2DEE2F60AF524989FAF5''. Thus, the complete video ID in the Youku peer CDN can be denoted as [$C_{CDN}, C_{peer}$]. With this video ID, a original user is able to request the same video content from both of the CDN and the P2P network separately.

	 Next, we present the workflows of video streaming as follows:
	 
	 $\rhd$ Streaming protocol in Youku: The Youku video CDN adopts the unencrypted HTTP protocol in both signaling exchange and video streaming, while the peer CDN control plane adopts the encrypted TCP-based protocol for signaling between the scheduling servers and agents. It adopts its proprietary UDP-based and TCP-based P2P protocol for the agents to exchange their data.
	 
	 $\rhd$ When the user starts to watch a videos, firstly, she downloads the beginning part of the chunk from the edge servers (step \textsf{1}, \textsf{2}); At the same time, she queries the scheduling server for a peer list generated based on the user' location and ISP information (step \textsf{3}). If suitable peers are detected, the user and selected peer routers attempt to establish connection with each other and delivery chunks (step \textsf{4}). Meanwhile, the connection with the edge servers still hold on; if the selected peers become slow or unreliable, the CDN edge server can cover this difference. Then the user experience does not suffer from this instability. 
	 
	 $\rhd$ For the content deployment on the peer routers, each peer router periodically obtains the prefetching chunk list (step \textsf{C}), and then downloads them one by one either from the CDN servers (step \textsf{D}, \textsf{E}) or from other peers (step \textsf{F}) based on the scheduling of the peer CDN controller.

\subsection{Content Information} \label{serviceprovision}

In the Youku peer CDN system, the contents cached in the peer routers are video chunks. Fig.~\ref{fig:chunkinfo} shows the basic information of contents cached in our testing routers, including the chunk duration and size, and these distributions for contents in different routers are consistent. We can find out most of them are $6$ min and the average chunk size is $17$MB (hence one Youku smartrouter with the $8$GB storage can cache about $430$ chunks). This chunk-caching strategy makes the chunks of the same content spread across multiple routers, which achieves the load balancing between routers to serve users. 

The video bitrate is another important factor for VoD service. We use the tool \textsf{ffmpeg} to obtain the chunks bitrates cached in the smartrouters, and observe $3$ types of video bitrates, i.e., $270$Kbps (Standard Definition), $600$Kbps (High Definition) and $1200$Kbps (Super Definition). The higher quality videos are generally partitioned to shorter chunks, which can benefit the fine-grained streaming scheduling.

In the next section, we perform in-depth strategy analysis deployed in the smartrouter-based peer CDN.

\section{Strategies Used in the Peer CDN} \label{section:strategy}

In this section, we investigate the content deployment and peer selection strategies in the video peer CDN, which can successfully meet the user demands and improve the QoS of peer routers.

\subsection{Traffic Pattern}

\begin{figure}[!t]
%\vspace{-0.3cm}
\centering
 \begin{minipage}[t]{0.8\linewidth}
    
    \includegraphics [width=1\linewidth, height=3cm]{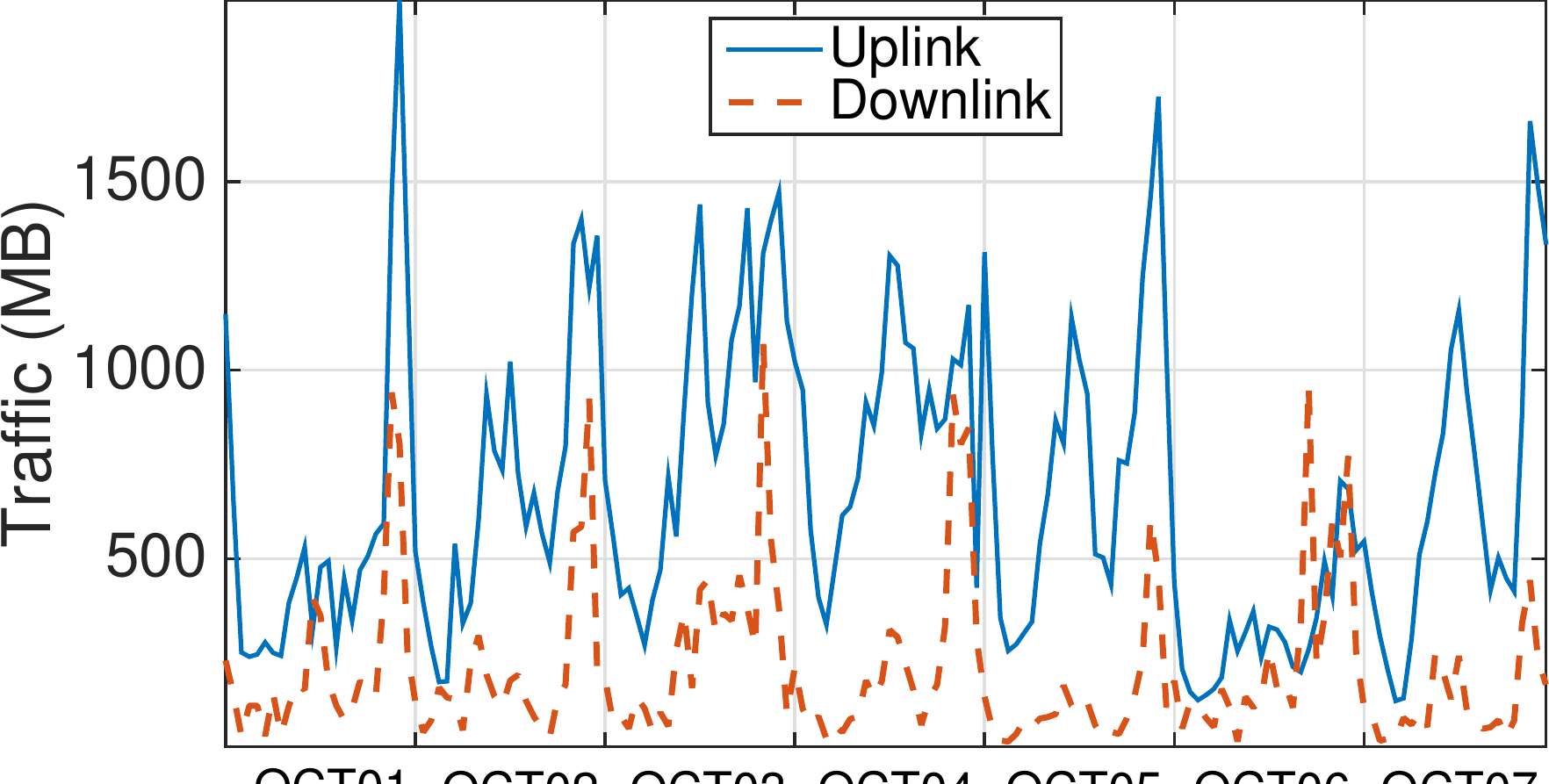} 
%    \centerline{\scriptsize (a) $R1$ traffic pattern.}
  \end{minipage}

  \caption{Traffic pattern of the peer router \textsf{R1}.}
  \label{fig:traffic_pattern}
  %\vspace{-0.2cm}
\end{figure}
\begin{table}[!t]
  \caption{The average amount of traffic delivered by testing routers in $1$ day. }
  \label{tab:5routers}
  \centering  
   %\begin{threeparttable}
  \begin{tabular}{lll}%{ *3{>{\centering\arraybackslash}m{1.5 in}} @{}m{0pt}@{}}
    \toprule
        Router	&	Upload traffic (GB)& Download traffic (GB)\\
    \midrule
        \textsf{R1}		&	 $19.90$ &   $6.10$ \\
        \textsf{R2}		&	$20.99$  & $6.23$\\
	    \textsf{R3}		&	$3.21$  &  $6.05$ \\
	    \textsf{R4}	    & $3.78$  &  $2.67$ \\
	    \textsf{R5}	  	&	 $22.03$  & $6.84$\\
     \bottomrule 
  \end{tabular}
   \vspace{-0.2cm}
\end{table}

\begin{figure*}[!t]
 \begin{minipage}[t]{0.32\linewidth} %0.215
    \centering
    \includegraphics [width=1\linewidth]{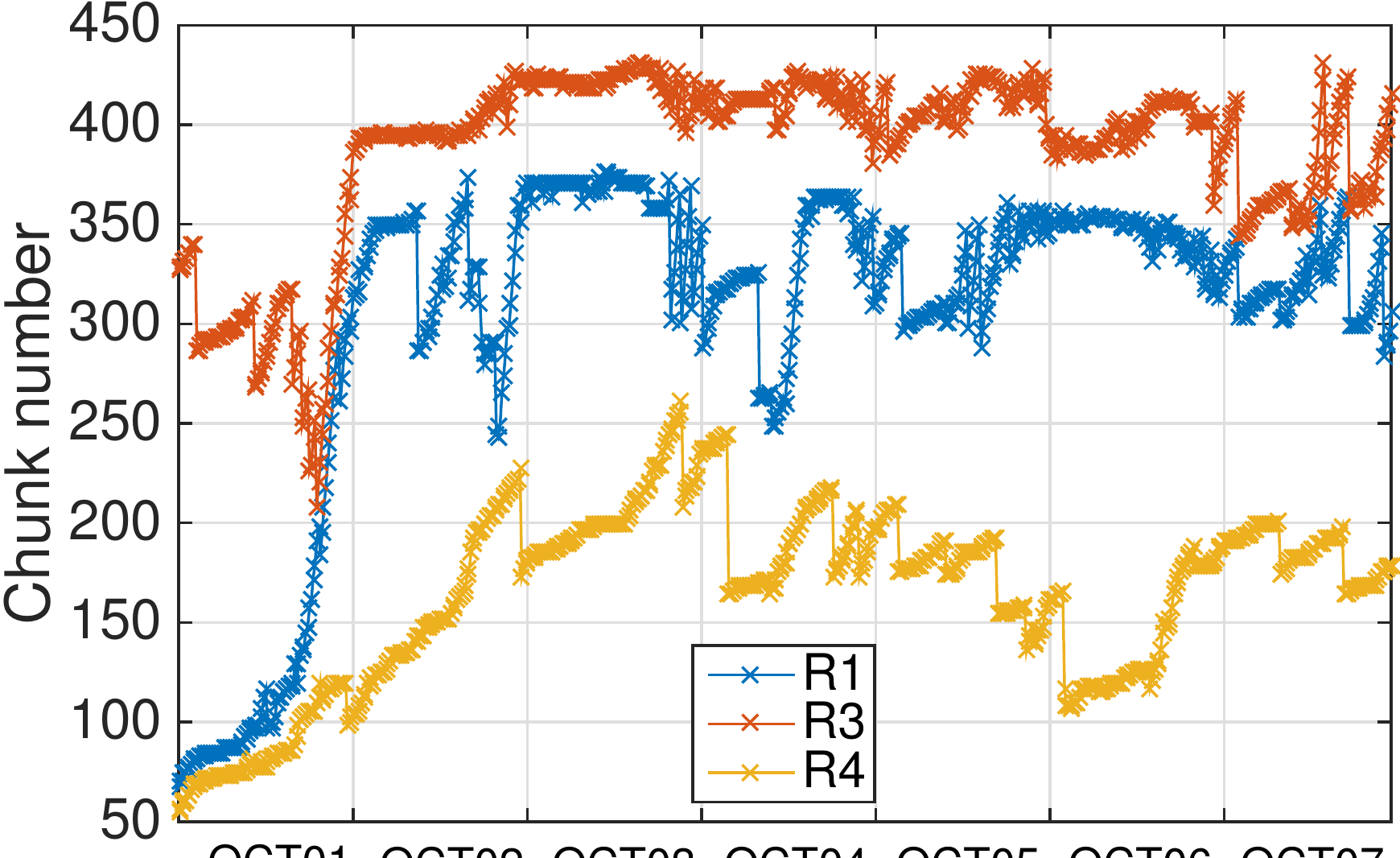}
		\centerline{\scriptsize (a) The change of the number of chunks cached in the routers.}
  \end{minipage}
  \hfill
\begin{minipage}[t]{0.32\linewidth} %0.25
    \centering
    \includegraphics [width=1\linewidth]{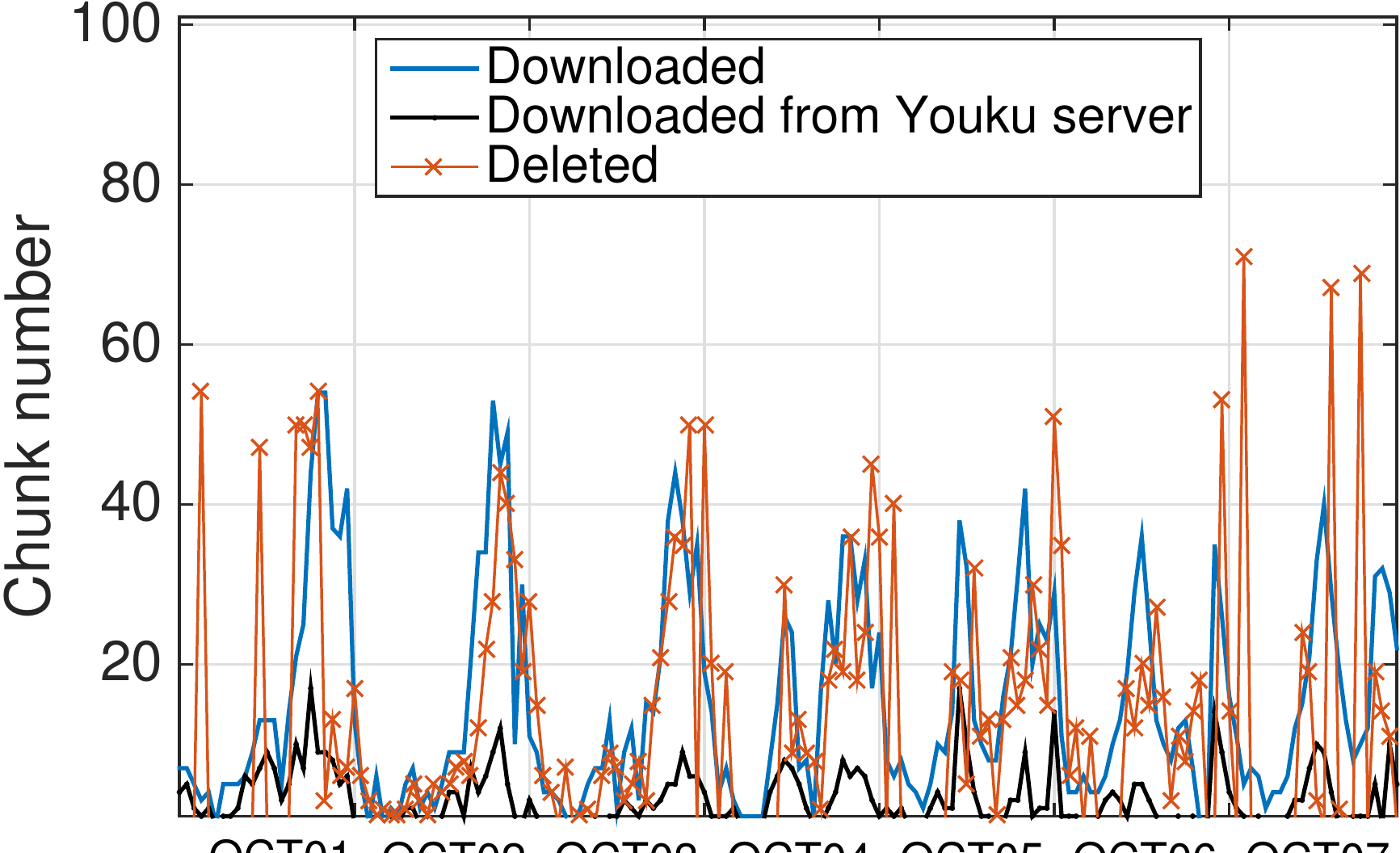}
		\centerline{\scriptsize (b) The number of hourly downloaded/removed chunks.}
  \end{minipage}
  \hfill
  \begin{minipage}[t]{0.31\linewidth} %0.25
    \centering
    \includegraphics [width=1\linewidth, height=3.5cm]{chunk_live_time.eps}
		\centerline{\scriptsize (c) The distribution of chunk lifespan.}
  \end{minipage}
  \hfill
  \caption{The characteristics of chunk replication scheduling.}
  \label{fig:chunknumber}
   \vspace{-0.2cm}
 \end{figure*}

 First, we overview the working condition of our $5$ peer routers. In Fig.~\ref{fig:traffic_pattern}, we depict the upload and download traffic of \textsf{R1} over $7$ days, and the other days and other routers exhibit the similar pattern. Our observations are as follows: (1) The volume of data uploaded by the router is larger than the volume of data downloaded, indicating that smartrouters serve as traffic ``amplifiers''; (2) Downlink traffic shows a periodical manner (exactly like the traffic pattern in the conventional video services \cite{yu2006understanding}), indicating that smartrouters can be scheduled to fetch content in a centralized manner.  

Our $5$ testing routers exhibit similar download and upload patterns. But on account of the different networks they are deployed in, their amount of download and upload data is different. Table~\ref{tab:5routers} shows the average amount of traffic delivered by the $5$ during one day. We observe \textsf{R1}, \textsf{R2} and \textsf{R5} have the comparable results, due to they are in the same ISP and have the same access bandwidth. For the \textsf{R2}, its upload capacity is restricted by the NAT traversal. As \textsf{R4} is deployed in the worst network condition within these $5$ routers, its upload/download capacity of \textsf{R3} is the at the minimum level.

Next, we will study the concrete replication strategies used for the peer routers. Note that our $5$ smartrouters show consistent caching strategy, thus we present the patterns of one router in this paper (unless otherwise specified).

\begin{figure}[!t]
      \begin{minipage}[t]{0.9\linewidth} %0.21
    \centering
   	\includegraphics [width=\linewidth]{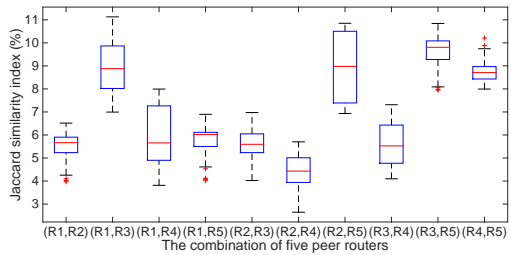}
   \end{minipage}
  \hfill
 \caption{Chunk similarity between the $5$ routers.}

  \label{fig:chunkscheduling}
   \vspace{-0.2cm}
 \end{figure}

\subsection{Content Deployment}\label{section:chunk}

  \begin{figure*}[!t]
\begin{minipage}[t]{0.95\linewidth} %0.25
    \centering
    \includegraphics[width=1\linewidth]{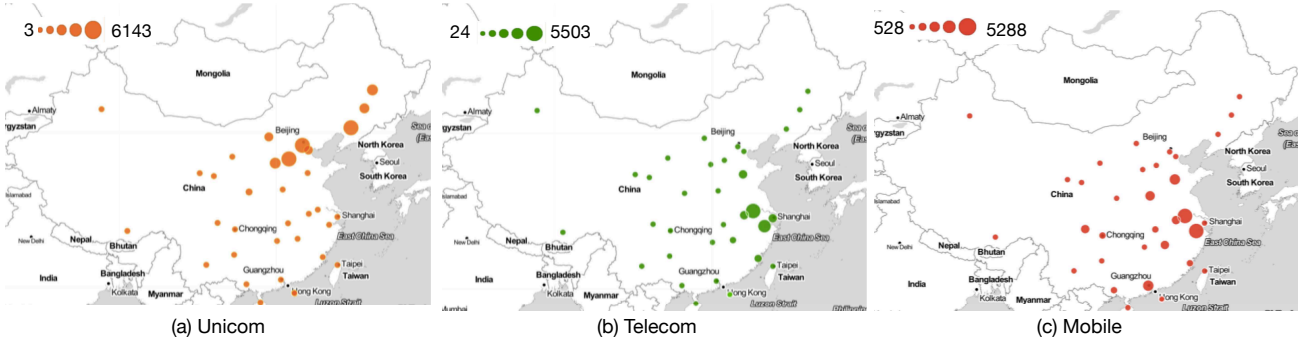}
  \end{minipage}
 \caption{The geo-distribution of peers in the different ISPs, including China Unicom, China Telecom and China Mobile.}
  \label{fig:peergeo_dis}
   \vspace{-0.2cm}
 \end{figure*} 
 
\subsubsection{Content Replication and Replacement}

We study the Youku replication strategy by analyzing how and when chunks are downloaded or deleted over time. Fig.~\ref{fig:chunknumber}(a) shows the change of the number of chunks cached in the testing routers \textsf{R1}, \textsf{R3} and \textsf{R4} over $1$ week (the following days are similar). We observe that the number of chunks increases gradually at the beginning of the first $2$ days, then remains nearly at a constant, i.e., although every testing router has a $8$GB internal TF card, \textsf{R1}, \textsf{R3} and \textsf{R5} keep about $427$, $370$ and $250$ chunks at most in their storage, respectively. Thus we conclude that the network condition is a crucial factor for Youku to decide the smartrouters' practical storage utilization ratio.

In Fig.~\ref{fig:chunknumber}(b), we plot the number of chunks downloaded to the \textsf{R1} and removed from the \textsf{R1} in each hour during $7$ days. The downloaded chunk number follows a daily pattern, and we can speculate that the Youku centralized controller invoke the peer routers to fetch chunks periodically in the form of fine grained timeslot (maybe hourly or shorter). Moreover, we count the number of chunks which are downloaded from the CDN edge servers. During this week, $22$\% of the chunks are downloaded from CDN servers. Thus most of the chunk deployments are finished by peer routers themselves, which effectively alleviates the load of Youku servers for content deployment. 

As for the chunk removal, it is also scheduled by the centralized control, instead of only using the cache replacement algorithms (e.g., Least Recently Used algorithm). For example, $55$ chunks are deleted in Oct 1 when there is spare storage capacity.

Another important question about the router caching is the chunk update. Fig.~\ref{fig:chunknumber}(c) plots the distribution of chunk lifespan. The median (reps. mean)  chunk lifespan is $24.2$ hours (resp. $29.8$ hours), i.e., most of the chunks cached in the peer router last about one day. There are plenty of fresh and popular published contents that original users want to watch, but the router storage is much smaller than the CDN edge server, which results in a frequently and timely contents update in the peer router. Furthermore, this rate of chunk update is similar to the change rate of top ranked videos measured in the VoD systems \cite{yu2006understanding, PPTVmobile}.

%We make the conclusion that chunks would be timed deleted, or deleted when the current storage capacity is not enough to complete the new incoming chunk download task. But if peer CDN system find this chunk is still popular, the replication scheduling server would demand the peer routers to download the chunk again.

In order to figure out whether Youku peer CDN makes a difference between the peer routers by their ISPs or regions when it conducts the content replication, we evaluate the chunk similarity between any two peer routers by using Jaccard index \cite{jaccard}, i.e., the size of the intersection of two chunk sets divided by the size of the union of two chunk sets. In Fig.~\ref{fig:chunkscheduling}, we calculate the similarity index every $5$min during our measurement period, which is depicted by a box-and-whisker diagram. It shows the average similarity indexes of any two peer routers are between $4.2$\% to $9.8$\%, which remains at the same degree, even if the peer routers in the different ISPs and locations, indicating the smartrouters are scheduled to cache videos with same opportunities.

There are two reasons that enforce the real-time replication. Firstly, increasingly hot videos are generated at anytime and anywhere, such as social news and user-generated content. Secondly, most of the copyrighted contents are enforced to release at prime television time every day. Thus content pre-deployment is a tough work, even if peer CDN providers can predict which video would be popular. 

 \emph{The peer CDN enables prompt global scheduling over millions of peer routers, including pushing newly published videos to peer routers and dynamically replacing staled contents at these routers. Such prompt and global strategies are enablers for today's frequently changed user interests.}

 \begin{figure}[!t]
\centering     
       \includegraphics[width=0.8\linewidth]{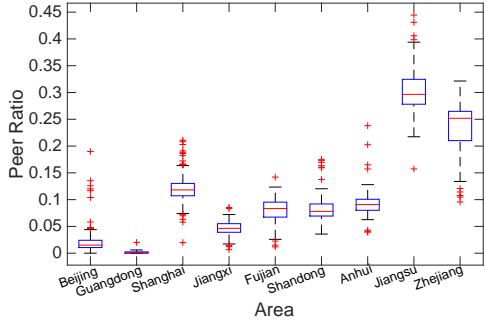}
	      \caption{The peer distribution of chunks in different locations.}
	      \label{fig:chunkpeer}
	       \vspace{-0.2cm}
\end{figure}

\subsubsection{Peer Router Management from a Global Perspective}

 A large amount of peer routers need an efficient management system for the content deployment and service delivery. In order to figure out the management mechanism and obtain the ``snapshot'' of the Youku peer CDN from a global perspective, we apply a crawler to track both of the peer geo-distribution and content geo-distribution in the Youku peer CDN.

We sample $70$ chunk cached in the testing routers, and design a crawler to collect the their peer lists by using these chunk IDs, i.e., the step \textsf{3} in the Table~\ref{tab:HTTP}. Consider that peers may join and leave the video swarm dynamically, we crawl the candidate peer lists of all the chunks about $30$min (it does not trigger the DoS attack alert from Youku). We conduct the crawling from $3$ dominating ISPs in Beijing and Shenzhen, including China Unicom, China Telecom and China Mobile. After combining the $6$ sets, We find unique $96,870$ peers in total.

From the results, we observe the peer list request is redirected to different scheduling servers based on where this request come from. In there, requests from Unicom, Telecom, Mobile are redirected to the scheduling servers which are also located in Unicom, Telecom, Mobile, respectively, and above $90$\% of the collected peers are located in the same ISP as the scheduling servers. Then we confirm that Youku peer CDN manages the pool of peers in a distributed mechanism according the ISP. We further map the IP addresses to the city-level location in the China mainland by querying the \textsf{nali}\footnote{https://github.com/meteoral/Nali} database. Fig.~\ref{fig:peergeo_dis} shows most of the Unicom peers are located in the northeast China, while most of the Telecom and Mobile peers are located in the Southeast China. This distribution follows the common sense of ISP deployment in China.

To explore the content deployment condition with a global view, using the peer dataset crawled from the Telecom, Fig.~\ref{fig:chunkpeer} plots a box-and-whisker diagram to show the peer geo-distribution of the $70$ chunks. We observe the peer distributions of different chunks with different popularities are similar, e.g., most chunks aggregate in Jiangsu and Zhejiang in accordance with Fig.~\ref{fig:peergeo_dis}(b), which further verifies that the chunk copies are deployed in the pool of routers with a global consistent mode.
 
\subsection{Peer Selection and Download Scheduling} \label{peerselection}

 \begin{figure}[!t]
\begin{minipage}[t]{0.9\linewidth} %0.25
    \centering
    \includegraphics[width=1\linewidth]{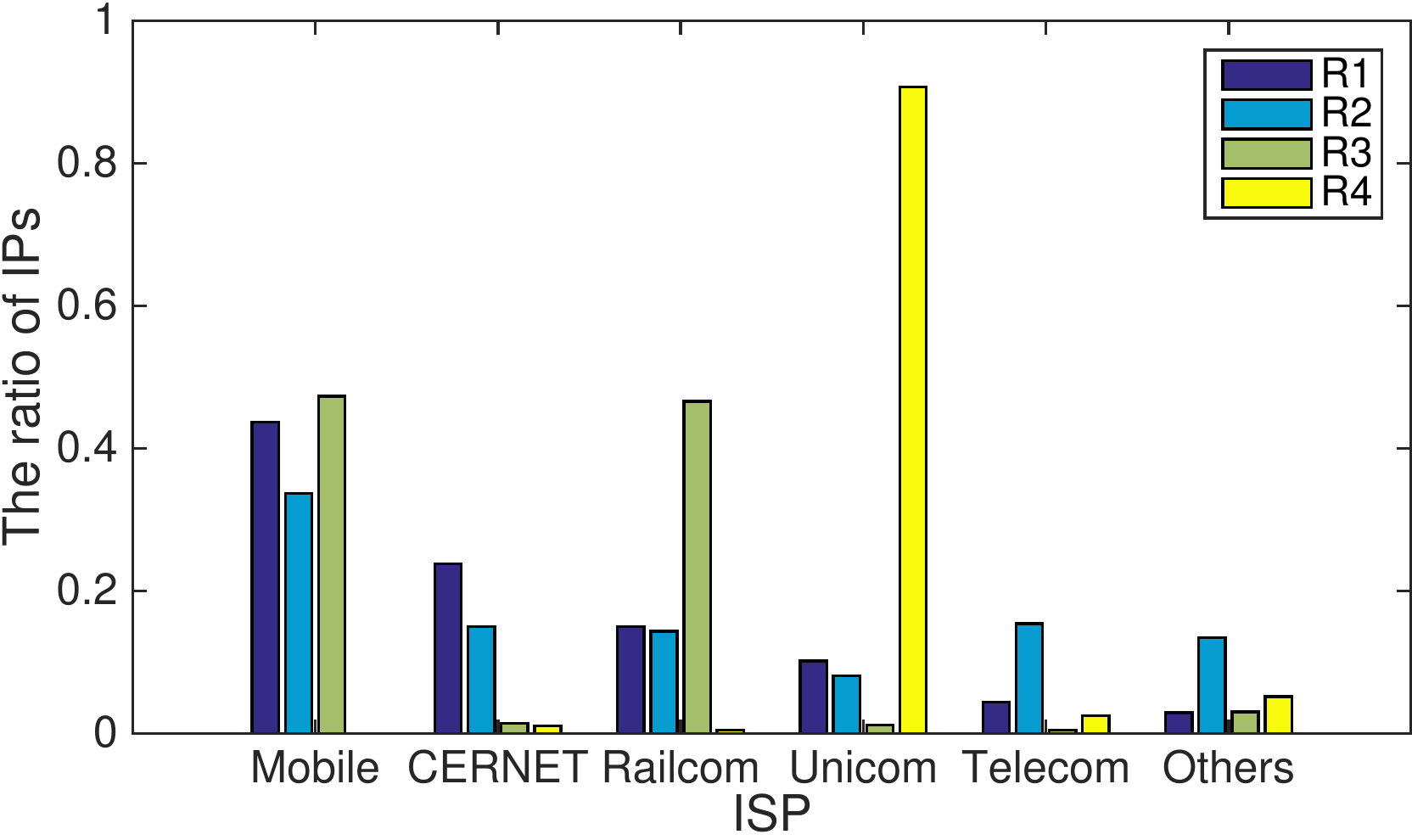}
		\centerline{\scriptsize (a) The ISP distribution of Top-100 IP.}
  \end{minipage}
  \\
  \\
  \hfill
  
     \begin{minipage}[t]{0.9\linewidth} %0.215
    \centering
    \includegraphics[width=1\linewidth]{IP_area_outall.eps}
		\centerline{\scriptsize (b) The Area distribution of Top-100 IPs.}
  \end{minipage}
    \hfill
 \caption{The distribution of Top-100 IPs in uplink traffic.}
  \label{fig:peer_dis}
   \vspace{-0.2cm}
 \end{figure} 
 
We further investigate the peer selection strategy adopted by Youku, i.e., how peer router chooses the partners for chunk uploading and downloading. For this purpose, we focus on the daily Top-$100$ IPs, which communicate with our peer router, ranked by the amount of traffic which they contribute during the peak time (e.g., $8$ PM -- $11$ PM) over the course of our measurement, and we analyze their distribution in terms of their ISPs and geographical locations.

 We select the top IPs of uplink traffic as the representative to analyze the peer selection. First, we plot the distribution of all selected Top-$100$ IPs in terms of ISP in Fig.~\ref{fig:peer_dis}(a). For the \textsf{R1} - \textsf{R3} routers deployed in the CERNET, although they are in the same region and ISP, they can join different peer sets to transfer chunks, e.g., \textsf{R1} and \textsf{R3} select most peers/users in the ISP Mobile, but \textsf{R2} uploads chunks mostly to the peers/users in the ISP Telecom. As we know, only the education organizations or research institutes can access to CERNET, which results in fewer routers deployed in CERNET and no significant ISP barrier between CERNET and other ISPs. Thus our peer routers can communicate with the peers from various other ISPs based on the global QoS monitoring redirection. As for \textsf{R4}, it mainly uploads chunks to Unicom peers ($89$\%) to avoid the ISP barrier problem.

The geographical location distribution of these Top-$100$ IPs is presented in Fig.~\ref{fig:peer_dis}(b). We observe that peers which download contents from \textsf{R1} - \textsf{R3} are located in different locations with relatively uniform ratio. While the \textsf{R4} mainly upload contents to the peers located in its nearby location, i.e., Beijing and Heibei province.

Based on the analysis above, we summarize our observations as follows: \emph{Such peer CDN can form an effective QoS monitoring sub-system. As peer routers are scheduled by a centralized peer selection mechanism using global knowledge, e.g., based on ISP or location, they can be assigned to serve users effectively. Thus large overall bandwidth can be achieved when peer routers are effectively matched to serve particular users.}

\section{The system performance} \label{section:qos}

In this section, we examine the system performance of the Youku peer CDN. The primary goals for the content providers to push content resources to the edge of network are: (1) Improve the service quality experienced by users through shortening the distance between users and content resources; (2) Alleviate the bandwidth occupancy costs of the CDN infrastructure by redirecting user requests to peer routers as more as possible. In this section, we evaluate the system performance upon these two targets.

\subsection{Performance Perceived by Smartrouters} \label{sec:routerqos}

First, we present the peer CDN performance using the traffics captured from the testing routers based on the QoS metrics, such as latency and download speed. We choose \textsf{R1} and \textsf{R4} as the representatives to demonstrate the results (The results of \textsf{R1}, \textsf{R2}, \textsf{R3} and \textsf{R5} are similar).

\begin{figure}[!t]
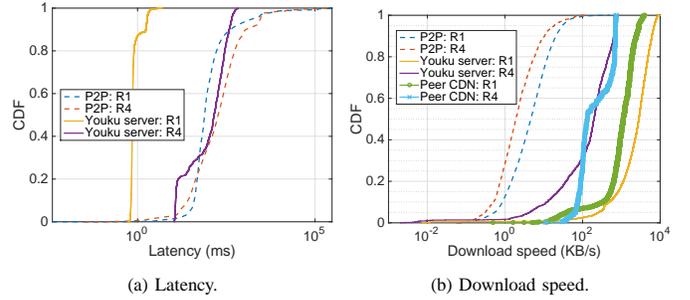

 \centering
 
 		\begin{minipage}[t]{0.48\linewidth}
 		\centering
 		\includegraphics[width=\linewidth]{stream_cdf25All.eps}
		\centerline{\scriptsize (a) Latency.}
 		\end{minipage}
 		 \hfill	
 		 \centering		
 		\begin{minipage}[t]{0.48\linewidth}
 		\centering
 		\includegraphics[width=\linewidth]{stream_cdf_throughputAll.eps}
 		\centerline{\scriptsize (b) Download speed.}
 		\end{minipage}
 		 \hfill	
 		 \caption{The QoS performance of \textsf{R1} and \textsf{R4}.}
 		 \label{fig:router_qos}
 		  \vspace{-0.2cm}
 \end{figure}

 In this paper, we denote the latency of a network connection as the duration from its first packet sent by the router to the next packet received by the router. As shown in Fig.~\ref{fig:router_qos}(a), for the \textsf{R1}, we observe there is an obvious gap between the P2P connections and \textsf{R1}--Server connections, i.e., the average latency for downloading contents from servers is about $1$ms, while the average latency for P2P communications is about $100$ms. But for the \textsf{R4}, there is no significant difference between the latency of P2P connections and the \textsf{R4}--Server connections, indicating that the bottleneck is the downlink of \textsf{R4} ($4$Mbps) which results in the long delay.
  
   Fig.~\ref{fig:router_qos}(b) compares the download speed of: 1) single P2P connections, 2) Router--Server connections, and 3) parallel P2P connections of downloading the same chunk, which is labelled as ``peer CDN''. From the results, we observe that the download speed of a single P2P connection is slower, but the total peer CDN speed is quite high which can come up with the speed of downloading contents from CDN servers.

From the analysis results of the smartrouters, we perceive that the peer video CDN should overcome the obstacles of impaired P2P connections, i.e., higher latency and lower download speed, to guarantee the QoS experienced by users.

\begin{figure*}[!t]
     \begin{minipage}[t]{0.285\linewidth}
          \centering
               \includegraphics[width=\linewidth]{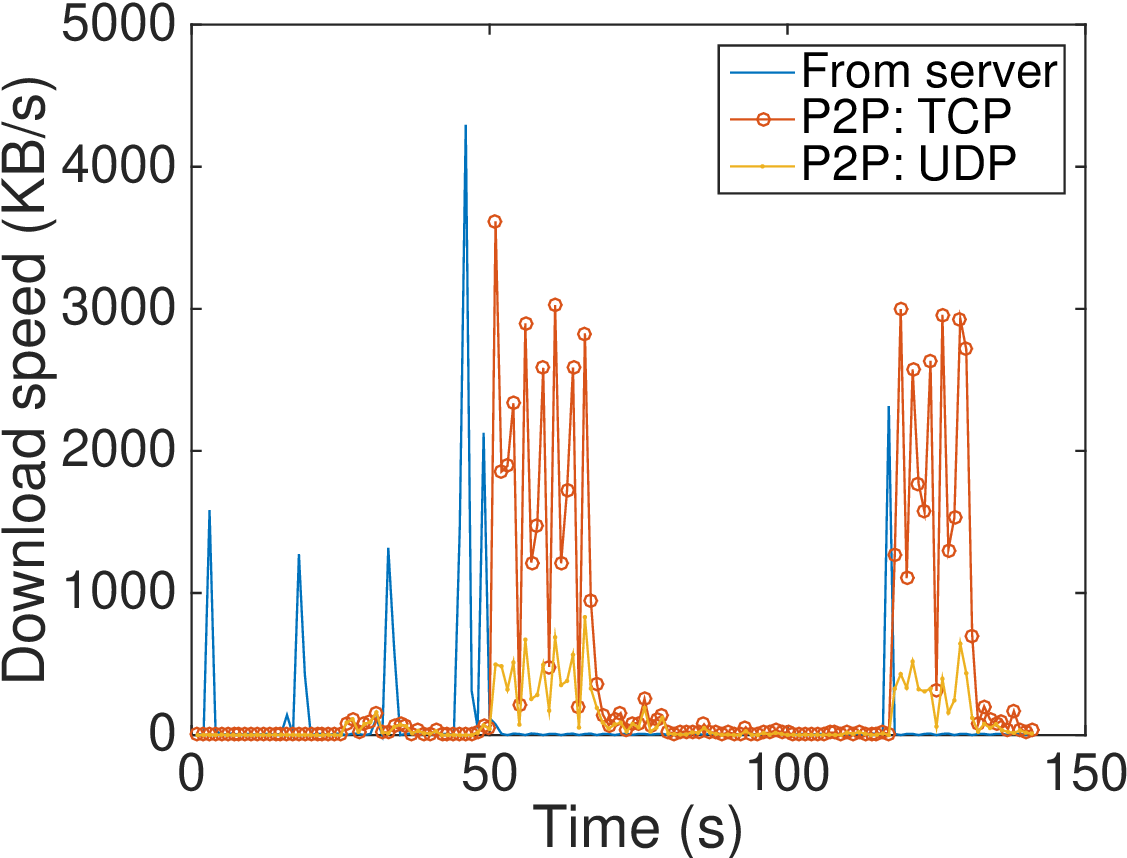}
	      \caption{The traffic pattern of a video session in the peer CDN.}
	      \label{fig:client_demo}
		% \vspace{-0.3cm}
     \end{minipage}
	\hfill
     \begin{minipage}[t]{0.63\linewidth}
		 			\begin{minipage}[t]{0.49\linewidth}
			  \centering
				   \includegraphics[width=\linewidth, height=3.9cm]{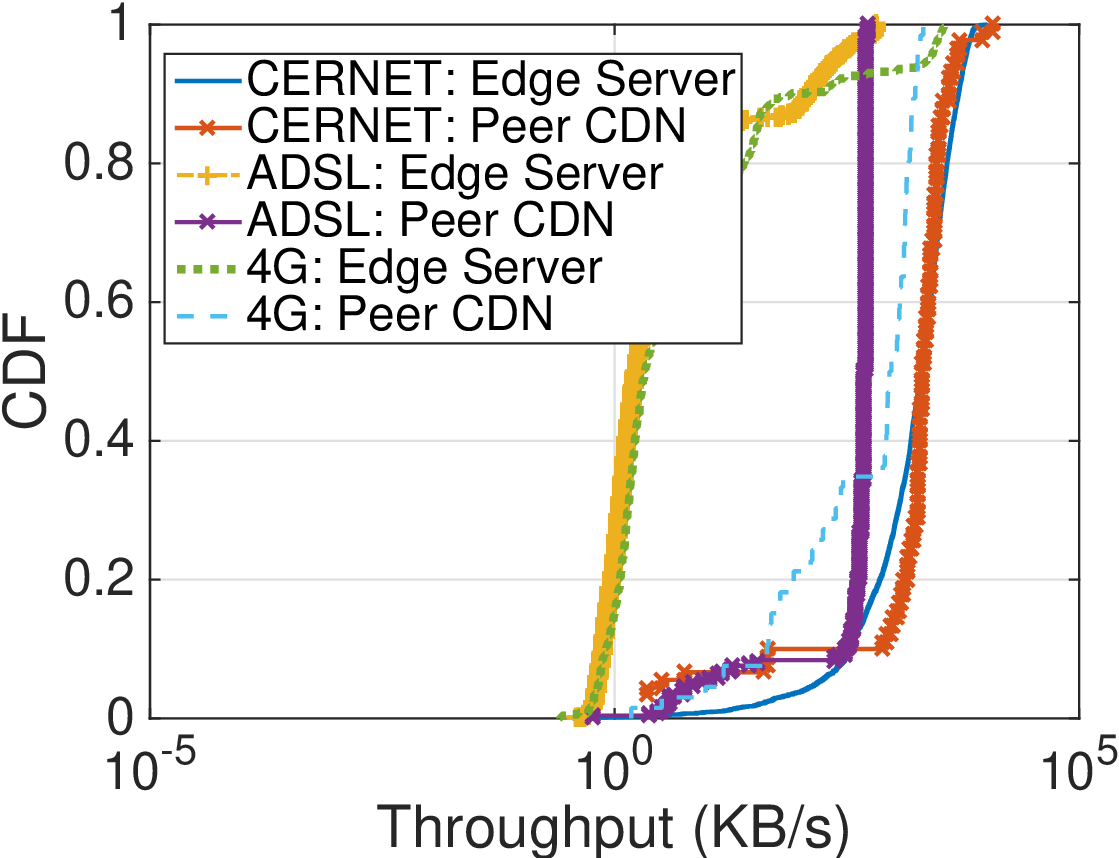}
			  \centerline{\scriptsize (a) Download speed from edge server vs. peer CDN.}
		 \end{minipage}
		 \hfill
		 \begin{minipage}[t]{0.49\linewidth}
			  \centering
				   \includegraphics[width=\linewidth, height=4cm]{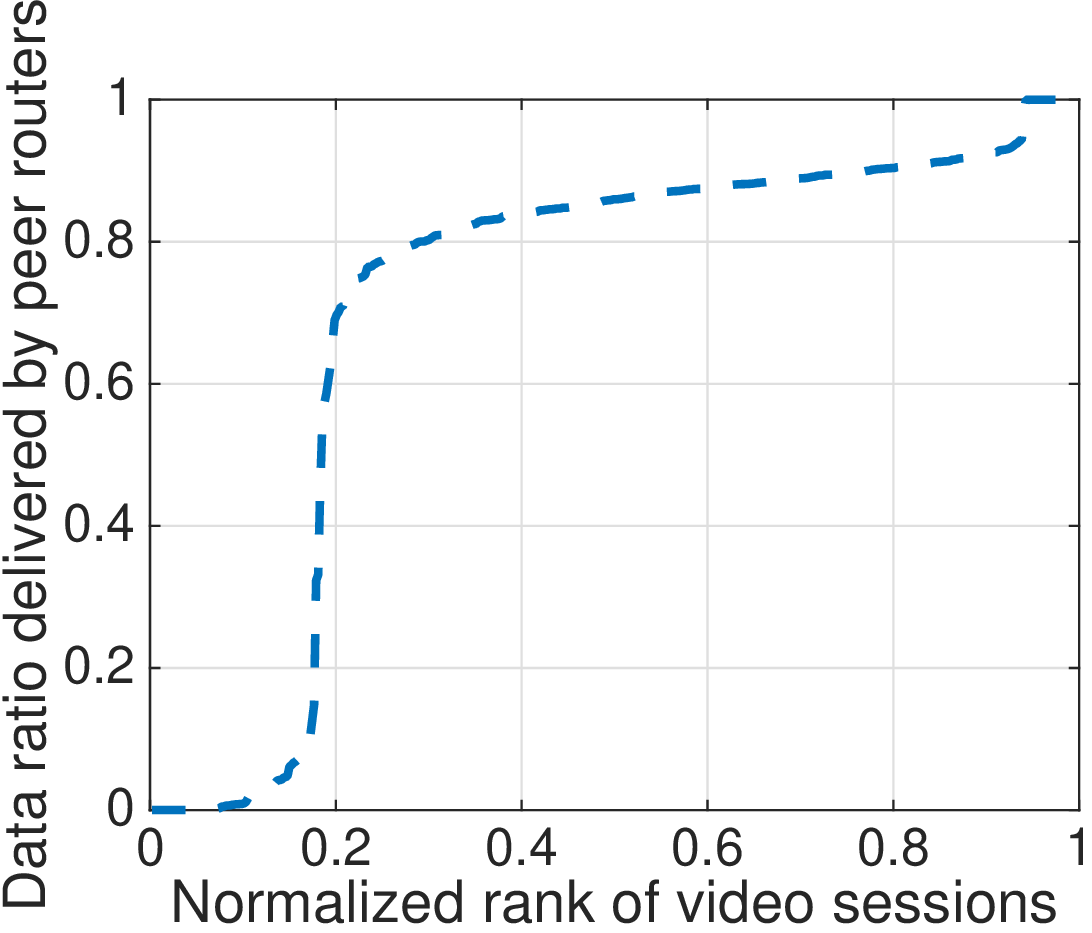}
			 \centerline{\scriptsize (b) Peer CDN efficiency.}
		 \end{minipage}
		 \hfill
		    \caption{The video delivery QoS of peer CDN perceived by users.}
     \label{fig:client_qos}
  %   client_p2pratio4g.eps
     % \vspace{-0.5cm}
   	\end{minipage} 
   	\hfill
   	     	\vspace{-0.2cm}
\end{figure*}

\subsection{Performance Perceived by Users}   \label{sec:userqos}

 In this part, we study the performance experienced by the original users in the peer CDN. 
 
 Fig.~\ref{fig:client_demo} illustrates the general traffic pattern of a VoD session. The users firstly requests the video from CDN edge servers and watches the beginning part of the video, which overcomes the start-up delay in the traditional P2P system demonstrated in the Fig.~\ref{fig:router_qos}(a). Then she downloads the chunk from multiple candidate peers in parallel, which guarantees the total download speed experienced by users.

 We evaluate the video delivery performance in Beijing with different network environment, including CERNAT ($80$ Mbps), Unicom ADSL ($4$ Mbps), and Unicom $4$G ($8$ Mbps). We choose $450$ videos on the Youku website and perform a series of video sessions based on their popularity given by its website\footnote{\url{http://index.youku.com/}}, which follows Zipf distribution. For each video session we record the average download speed and data ratio coming from peers. Fig.~\ref{fig:client_qos}(a) compares the chunk download speed of 1) peer routers transport bytes which larger than 60\% of the chunk, and 2) all bytes are downloaded from CDN edge servers. We can observe that the peer CDN has stable download speed in our experiment. Even in the Unicom ADSL and Unicom $4$G, peer CDN achieves higher download speed compared to that from edge servers, the reason is that multiple peer routers parallelly accelerate the speed (there are about up to $20$ concurrent connections to download a chunk in our experiments). 
 
 Then we evaluate the peer router delivery ratio and analyze which types of videos can be delivery by peers. Fig.~\ref{fig:client_qos}(b) shows the data ratio delivered by peer routers in our datasets. We observe $80\%$ of the content requests can be served by peer CDN with at least $70$\% of the bytes came from peer routers. Most of the videos with lower peer delivering ratio are user generated content (UGC) which are out-of-date. It is worth noting that the out-of-date UGC ``Gangnam Style'', which was the most popular videos in many video websites, can keep up with $66\%$ data ratio from peers.
  
  In summary, Youku peer CDN delivery QoS is comparable or better to that of the CDN system; and a substantial fraction of the content requests can be served by the peers.  
  
  %We can simply calculate as follows: assuming the duration (resp. bitrate) of a movie is $2$h (resp. $1200$Kbps). If each movie have $10$K copies, $300$K Youku peer routers can cache $233$ different movies, which can meet the interests of most users and consequently alleviate the resource renting costs of the conventional CDN. It demonstrates network resources of end-user have great potential to provide great VoD service, by utilizing the smartrouter storage and idle uplink bandwidth. 

\section{Discussion} \label{sec:discussion}
In this section, we discuss the limitation of the peer CDN and propose two potential principles which can be influential for peer CDN design.

$\rhd$ Limitation: In order to effectively serve users, massive amount of videos are continuously replicated between peer routers, which consumes too much user-side bandwidth. In our measurement, distinguishing which contents are uploaded to end users and  which contents are uploaded to other peer routers is difficult. But it is worthy to verify this problem and compare the traffic consumption with the performance gain.

$\rhd$ Potential improvement: 1) Global QoS monitoring based on Peer routers: Through the traffic datasets analysis, we observe that a large amount of flows between peer routers maintain the long time but light load connections. We speculate that there are probing flows to make peer routers keep contact with their partners. Since the CDN control plane can obtain the end-to-end QoS information from the large scale peer router swarm, it would be influential for the entire network perception and fine-grained user request redirection. 2) Regional content replication: In our experiment, videos are replicated in a coordinative manner. However, the content popularities of different regions are diverse\footnote{https://www.youtube.com/trendsmap} and influenced by the social networks \cite{zhi-acmmm2012, zhi-tmm-cpcdn2014}. If we want peer routers to provide a efficient nearby service, balancing the regional content popularity and global content popularity is important for the replication strategy design in peer CDN. 

Based on our analysis, compared to the conventional CDN which manages the QoS between edge servers and end users and updates the contents for a large region, tapping the internal potential of peer CDN is promising to provide more efficient video service.

\section{Related work} \label{section:relatedwork}

\subsection{Content Distribution Systems}
 Video content constitutes a dominant fraction of online entertainment traffic today. In current video service platforms, CDN and P2P are two representative techniques \cite{baochun-tomccap-streaming2013}. There are abundant measurement studies of content-distribution systems, including both of CDN-based VoD systems (such as YouTube \cite{adhikari2012youtube}, Netflix \cite{adhikari2012netflix} and Hulu \cite{adhikari2012hulu}), and P2P-based VoD systems (such as PPLive \cite{huang2008challenges} and Joost \cite{lei2010Joost}). From these measurement, both academia and industrial communities are aware that CDNs have the strong global controllability but weak scalability \cite{ZhanghuiControlPlane}, while P2P gets a strong flexibility but weak QoS guarantee for video delivery \cite{huang2008challenges}. 
	
\subsection{Peer CDN System}
	Recent years witness the ever-increasing amount of video traffic, e.g., Netflix and Youtube account for $55$\% of the downstream traffic with fixed access in North America by Dec 2015 \cite{Sandvine}, and Cisco predicts that over $3/4$ of the world mobile data traffic will be video traffic by 2020 \cite{cisco}. Some previous works have examined the peer CDN which combines the CDN and P2P system, which can obtain great benefit for content delivery. So far there are two main solutions to design peer CDN for video streaming, i.e., client-based strategy and smartrouter-based strategy. Client-based method urges clients download contents from each other when they cache the same contents, such as specialized applications including NetSession \cite{Akamai}, LiveSky \cite{YinLiveSky}, 3DTI \cite{arefin20124d} and PeerCDN \cite{wu2008peercdn}, and some web browser plug-ins \cite{FlowerCDN, Maygh}. All of these works do not involve the content prefetch strategies.
	
	Many researchers realize that the abundant resources, such as set-top \cite{cdn_orchestrate}, small cell base stations (SBS) \cite{poularakis2014approximation}, and Wi-Fi access points \cite{OfflineDownloading}, can be well utilized to assist the content delivery and offload the traffic of original server. These works provide theoretical content replication strategies. In this paper, we are interested to know how such design perform in the wild.
	
	Smartrouter-based peer CDN system has appeared for almost two years. \cite{thunder, zhang2015XunleiKanKan} study a router-based peer CDN system, Thunder, which is most like our paper. Their content scheduling strategy is very simple, i.e., push $80$TB traffic per day based on the file popularity in last day. \cite{Ming-nossdav2016} is our previous work about the content replication strategies used by the Youku peer CDN, which does not involve the detailed system analysis, such as the transport protocols, peer selection strategy and QoS performance.
	
	To the best of our knowledge, we are the first to use measurement to study the architecture, system strategies and performance of the real-world smartrouter-based peer CDN for VoD.

\section{Conclusion} \label{section:conclusion}

Smartrouter-based peer CDN starts a new CDN ecosystem, which deploys content delivery infrastructure (smartrouters) to the edge user sides, and leverages backhaul network resources contributed by end users. In order to understand the strategies, performance, limitation and potential impact on such content delivery system, in this paper, we conduct a comprehensive measurement on a real smartrouter-based peer CDN platform, deployed by ChinaCache and Youku, which serves $200$ million users of Youku. By passively and actively measuring the Youku peer router in different ISPs and locations, we provide the insights which are important for peer CDN. First, smartrouter based peer CDN system adopts a global replication and caching strategies. The cached contents are frequently updated on an hourly basis, in order to keep updated with the change of the content popularity. Second, such peer CDN deployment can itself form an effective QoS monitoring sub-system, which can be used for fine-grained user request redirection. Third, such peer CDN deployment can successfully guarantee the video delivery QoS, e.g., $80\%$ of the content requests can be served by nearby peer nodes. Finally, we discuss the system limitations and propose two potential design schemes, i.e., global end-to-end network monitor and regional content replication. 

%\section*{Acknowledgement}

%This work is supported in part by the National Natural Science Foundation of China (NSFC) under Grant No.61272231, 61133008 and 61402247, Beijing Key Laboratory of Networked Multimedia, and the SZSTI under Grant No.~JCYJ20140417115840259. The corresponding author is Lifeng Sun.

%


{
%\small
%\footnotesize
%\addtolength{\itemsep}{-6ex}
\bibliographystyle{IEEEbib}
\bibliography{mylib}
}

\end{document}